\documentclass[conference]{IEEEtran}
\IEEEoverridecommandlockouts

\usepackage{amsmath,amssymb,amsfonts}
\usepackage{algorithmic}
\usepackage{graphicx}
\usepackage{textcomp}
\usepackage{xcolor}
\usepackage[nolist]{acronym}
\usepackage[nospace]{cite}

\usepackage[acronym]{glossaries}

\def\BibTeX{{\rm B\kern-.05em{\sc i\kern-.025em b}\kern-.08em
    T\kern-.1667em\lower.7ex\hbox{E}\kern-.125emX}}
\begin{document}

\title{Evaluation of Deep Learning-based prediction models in Microgrids 
\thanks{This research was supported by the project EWIMA (project number:
	EFRE-0800681) funded by the European Regional Development Fund and
	Ministry of Economic Affairs and Energy of the State of North RhineWestphalia in Germany.}
} 

\author{\IEEEauthorblockN{1\textsuperscript{st} Alexey Gy\"ori}
\IEEEauthorblockA{\textit{Department of Information Management} \\
\textit{FIR e. V. an der RWTH Aachen}\\
Aachen, Germany\\
gy@fir.rwth-aachen.de}
\and
\IEEEauthorblockN{2\textsuperscript{nd} Mathis Niederau}
\IEEEauthorblockA{\textit{Department of Information Management} \\
\textit{FIR e. V. an der RWTH Aachen}\\
Aachen, Germany\\
ni@fir.rwth-aachen.de}
\and
\IEEEauthorblockN{3\textsuperscript{rd} Violett Zeller}
\IEEEauthorblockA{\textit{Department of Information Management} \\
	\textit{FIR e. V. an der RWTH Aachen}\\
	Aachen, Germany\\
	ze@fir.rwth-aachen.de}
\and
\IEEEauthorblockN{4\textsuperscript{th} Volker Stich}
	\textit{FIR e. V. an der RWTH Aachen}\\
	Aachen, Germany\\
	st@fir.rwth-aachen.de}

\maketitle

\begin{abstract}
It is crucial today that economies harness renewable energies and integrate them into the existing grid. 
Conventionally, energy has been generated based on forecasts of peak and low demands. 
Renewable energy can neither be produced on demand nor stored efficiently. 

Thus, the aim of this paper is to evaluate Deep Learning-based forecasts of energy consumption to align energy consumption with renewable energy production. 
Using a dataset from a use-case related to landfill leachate management, multiple prediction models were used to forecast energy demand.
The results were validated based on the same dataset from the recycling industry. 

Shallow models showed the lowest \ac{MAPE}, significantly outperforming a persistence baseline for both, long-term (30 days), mid-term (7 days) and short-term (1 day) forecasts. A potential decrease of up to 23\% in peak energy demand was found that could lead to a reduction of 3,091 kg in $\mathrm{CO}_\mathrm{2}$-emissions per year. 

Our approach requires low finanacial investments for energy-management hardware, making it suitable for usage in \acp{SME}.
\end{abstract}

\begin{IEEEkeywords}
Microgrids, Deep Learning, Renewable Energy Integration, Waste
Management, Artificial Neural Networks, Regression, Peak flattening, Nonlinear Optimization
\end{IEEEkeywords}

\section{Introduction}\label{intro}

The transition to renewable energy is still a major challenge for economies \cite{Tegart.1990,Pachauri.2015}. Countries differ in their ambitions to tackle the climate crisis and in their achievements while doing so\cite{PROGRAMME.2019}. By 2017, Germany reduced its $\mathrm{CO}_\mathrm{2}$-emissions compared to 1990 by 28\%,  while striving for an overall reduction of 40\% until 2020. While this goal will presumably be missed, the urgency to transform the economy to a more sustainable form increases. \cite{BundesministeriumfurUmweltNaturschutzundnukleareSicherheit.Juni2018}

While the transition to renewable energies reduces $\mathrm{CO}_\mathrm{2}$-emissions, new challenges arise \cite{Jacobsson.2006}.
The generation of most types of renewable energy depends strongly on meteorological factors such as wind and sun, making these resources highly unpredictable and uncontrollable \cite{Papaefthymiou.2016}. Hence, by increasing the share of renewable energy in the energy mix, the volatility in the electrical grid increases. This leads to increased stress on the existing electrical grid, which is designed to transfer energy from centralized fossil power plants to distributed energy consumers. \cite{Mwasilu.2014} 


One approach to address these issues is microgrids. Microgrids were introduced by Lasseter as a solution to integrate \acp{DER} into existing energy systems. There is no commonly accepted definition for the term Microgrid. There is a common understanding that "a Microgrid can be described as a cluster of loads, \ac{DG} units and \acp{ESS} operated in coordination to reliably supply electricity, connected to the host power system at the distribution level at a single point of connection, the \ac{PCC}." \cite{Olivares.2014}
As the overarching goal in all electrical grids is the sufficient quality of available power \cite{Gungor.2013}, the energy production needs to match the energy consumption at all times. This can be a highly complex task when taking into account peaks in energy demand, times when the Microgrid's energy production cannot meet its own demand, and the global distribution network has to be used.

Considering the aforementioned unpredicatability of renewable energy sources, other means must be considered to match energy production and consumption\cite{Hatziargyriou.2731Jan.2002}. One approach is \ac{DSM}.\cite{Strbac.2008} The \ac{dena} defines \ac{DSM} as "load management that is carried out on the basis of inter-company, energy-economical or grid-side incentives such as control power calls or price peaks on the electricity spot market."\cite{Seidl.062016}

In practice, this goal is achieved using load balancing\cite{Seidl.062016}. Load balancing can occur in the form of \textit{valley filling }and \textit{load shifting}. Flexible loads are activated, when the energy production is high and deactivated if energy production is low and the energy is used for more fundamental tasks. 

In this paper, the potentials of six prediction methods are investigated to predict loads. These predictions are used to balance loads and align energy production and consumption.

The paper is structured as followed. Chapter~\ref{Case} describes the case from which the dataset was obtained, chapter~\ref{Methods} describes the data preprocessing, forecasting methodology, and energy calculation used, chapter~\ref{Results} the results of the comparison, and chapter~\ref{Discussion} discusses the results. Chapter~\ref{Conclusion} concludes the paper and gives an outlook. 

\section{Case study}\label{Case}
\begin{figure}
	\includegraphics[width=\linewidth]{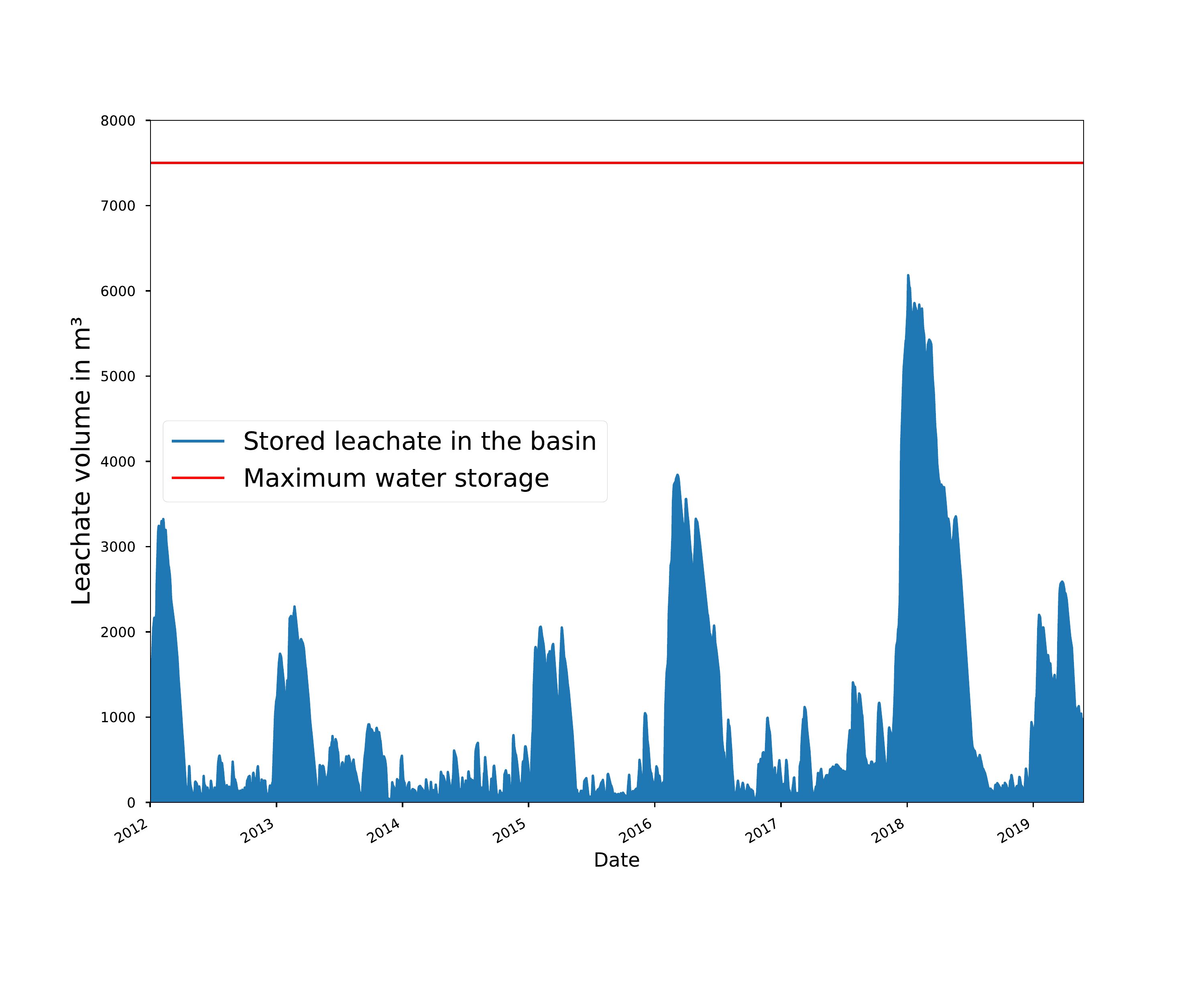}
	\caption{The current leachate in the reservoir basin over a period of 2704 days. The red line represents the basin's total capacity at 7500 m³.}
	\label{filling}
\end{figure}

This paper focuses on the use of energy forecasts at an open landfill as part of a waste treatment facility microgrid. Open, in this case, means that the landfill is not covered and rainwater can flow freely into the landfill. As the resulting leachate must not be allowed to pollute the groundwater, the leachate has to be collected in a basin and pumped off to a treatment facility \cite{S.Renou.2008}. This first step also acts as a part of the treatment; hence, the basin must always have a minimal amount of water inside due to the microorganisms requiring moisture. The historical water level inside the basin is depicted in Fig.~\ref{filling}. 

The water pumps operating on the leachate inside the basin pose a flexible load usable for peak shifting, as long as a minimum and maximum threshold of the volume of the basin is not reached. Additionally, the maximum output flow and input flow are not equal at 260 m³/d and 800 m³/d, respectively, making worst-case situations possible where overflow cannot be prevented. Against these requirements stands the potential of peak shifting of energy demand via \ac{DSM} which motivates creating plans via using forecasts of the future leachate flowing into the basin.
This paper evaluates the forecasting quality using state-of-the-art methods and, sequentially, analyzes scenarios where the forecasts can be used in order to shift energy demand to times where renewable energy is available in order to minimize peak loads. This has the potential to save on $\mathrm{CO}_\mathrm{2}$-emissions and increase the microgrid autonomy. The following scenarios will be used to evaluate the saving potentials:

\begin{enumerate}\label{scenarios}
	\item Usage of long-term forecasts (thirty days) to predict periods of high water inflow to minimize the risk of an overflow of the treatment facility due to extreme weather conditions.
	
	\item Usage of mid-term forecasts (seven days) to assist the holistic energy planning of the landfill in order to use available micro-grid energy from neighboring grid nodes.
	
	\item Usage of short-term forecasts (single day) to schedule short-term energy-intensive work at the landfill, e.g., repairs.
\end{enumerate}

In the given Microgrid, there are several renewable \ac{DG}, including a solar power plant and a combined power and heat power plant using locally collected landfill gas.

\section{Methods}\label{Methods}
\subsection{Data preprocessing}
\begin{figure}
	\includegraphics[width=\linewidth]{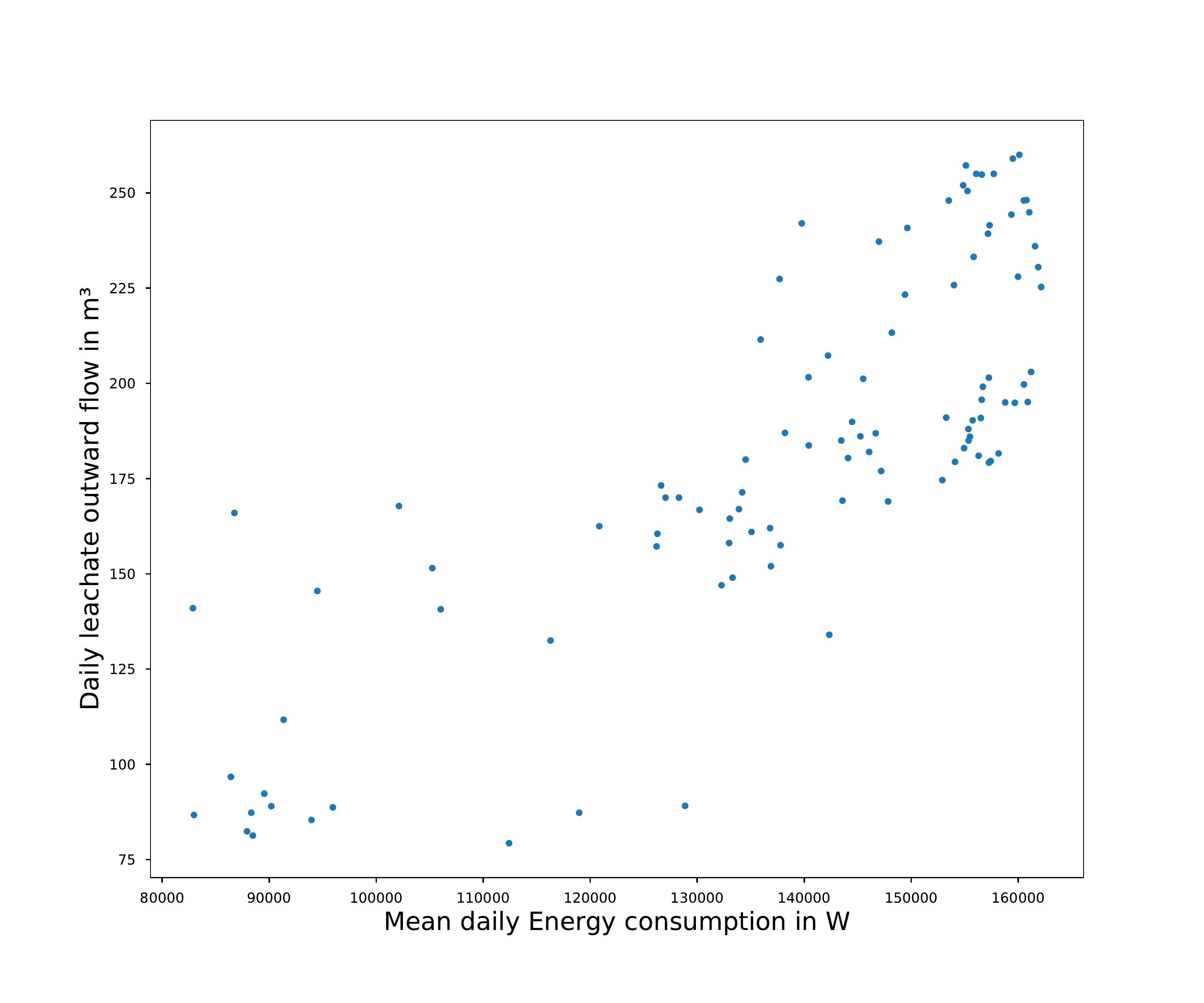}
	\caption{Scatter plot of the daily leachate pump output and the total energy consumption of the leachate treatment facility.}
	\label{energy_flow_scatter}
\end{figure}

Due to unavailable data on the specific factor of energy consumption of the pump depending on the pumping volume in m³, the relationship of the total energy consumption and the pumping volume in the leachate treatment facility was estimated using the Pearson correlation coefficient to analyze the linear relationship between both parameters for the available data of energy consumption for six months as depicted in Fig.~\ref{energy_flow_scatter}, resulted in a coefficient of $0.81$ which indicated a strong linear correlation.

Following this, we interpolated the pump energy consumption function of the leachate pump using an \ac{OLS} regressor resulting in the following estimate of the energy consumption in Wh for a single day: 

\begin{equation}
C(x_i) = 24h \times (388.12 \frac{W}{m^3} x + 67504.55 W)
\end{equation}
where $x$ is the pumping volume in $m^3$ on day $x_i$.

In addition to the given landfill water intake, exogeneous data on the landfill weather (rainfall, air pressure, temperature, wind direction, humidity) and the power supply of the landfill (battery voltage, power flow) was used in the prediction models. Since this data was only available for 635 days, this reduced the total time-series length. These different datasets are depicted in Table~\ref{datasets}.

\begin{figure}
	\includegraphics[width=\linewidth]{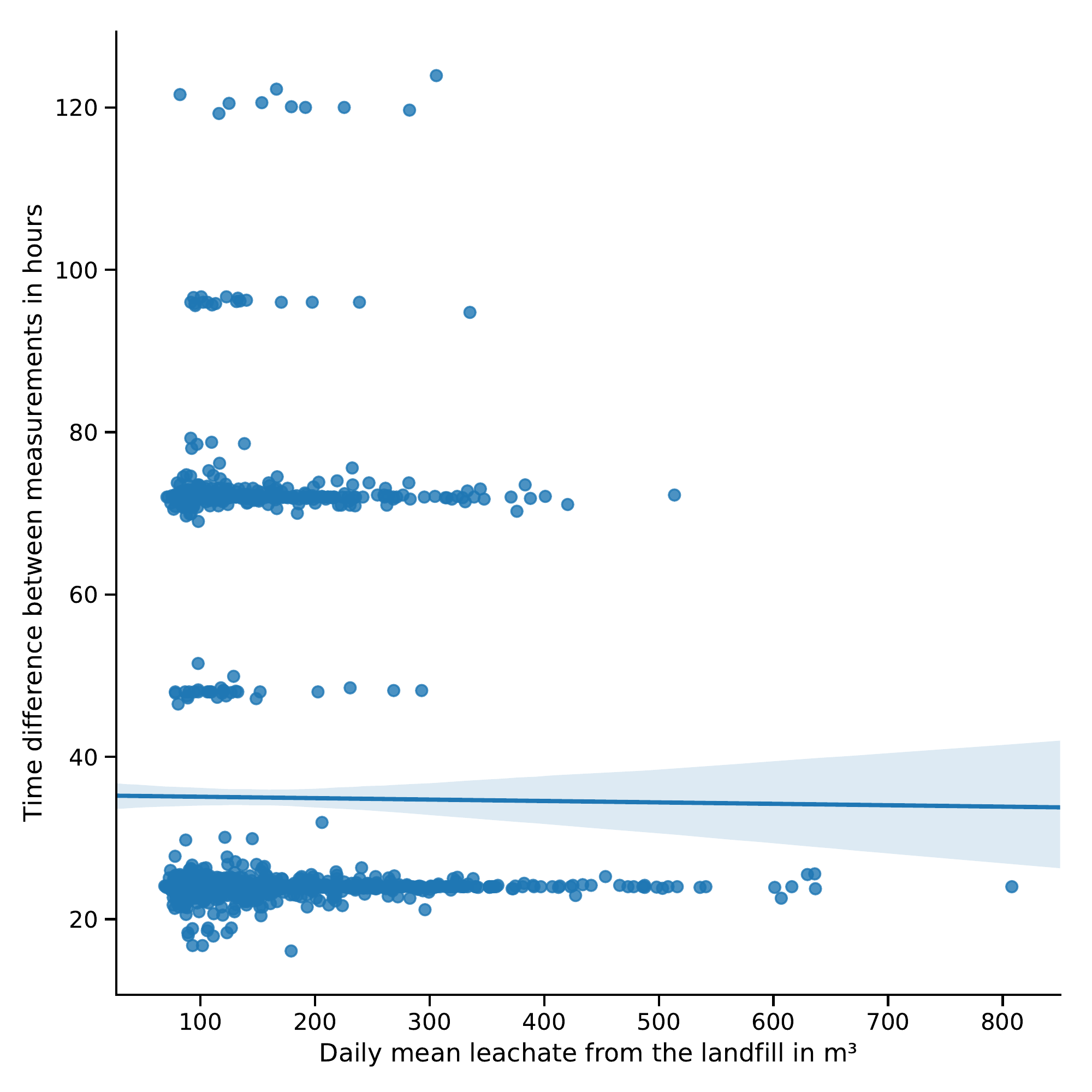}
	\caption{Scatter plot of the time difference from a single measurement to the subsequent measurement.}
	\label{time_difference_scatter}
\end{figure}

\begin{table}[htbp]
	\caption{The dataset configuration used for the forecast evaluation.}
	\begin{center}
		\begin{tabular}{|c|c|c|}
			\hline
			\textbf{Dataset} & \textbf{\textit{Data included}}& \textbf{\textit{num samples}} \\
			\hline
			inflow& target & 2704 days \\
			
			Merged data &  target + exogeneous & 635 days \\
			\hline
		\end{tabular}
		\label{datasets}
	\end{center}
\end{table}
As measurements were not done on weekends, the daily flow was linearly interpolated between weekdays. The plot in Fig.~\ref{time_difference_scatter} shows that the difference between measurements has little to no influence on the flow.

\subsection{Forecasting model specification}\label{forecasting_models}

This paper uses the following forecasting methods in comparison against the dataset:
\begin{enumerate}
	\item Baseline na\"ive persistence forecast using the mean of $k$ previous values. $\hat{y}_n = \frac{1}{k} \sum_{i=1}^{k}y_{t-i} $, 
	note that the forecast for all $n$ steps is equal.
	\item \ac{ARIMA} as the state-of-the art method for time-series forecasting. The AutoARIMA package was used to find the optimal parameters \cite{Smith.2019}.
	\item \acf{OLS} regression. 
	\item Dual-Stage Attention-Based \ac{RNN}, a state-of-the-art algorithm in forecasting \cite{Qin.472017}.
	\item Gradient tree boosting using the LightGBM library \cite{ke2017lightgbm}.
	\item Multi-layer \ac{ANN}. 
\end{enumerate}
The last three forecasting methods used a reformulation of the forecasting problem into a supervised learning problem via treating each time-step independently and providing the last $K$ values as input while forecasting the next $N$ time-steps.

\subsection{Forecast evaluation}
\begin{figure}
	\includegraphics[width=\linewidth]{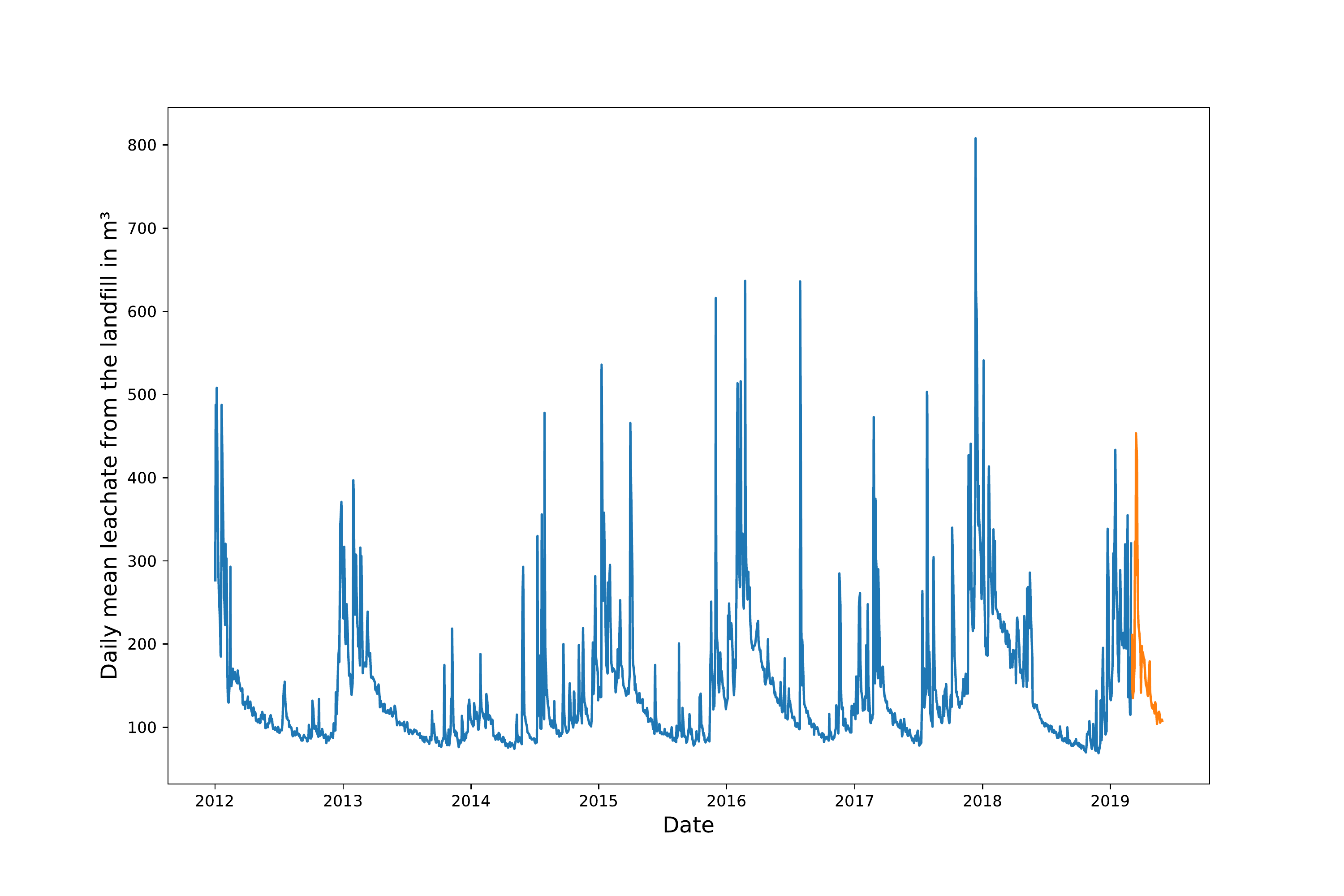}
	\caption{The total dataset of 2704 days of data}
	\label{train_test_split}
\end{figure}

The dataset is split into data reserved for validation and testing data. The split is shown in Fig.~\ref{train_test_split}, showing that the final 100 days of the available data was used for validation, while the rest was used for training. The chosen 100 days include a period of both high and low inflow variance. This relatively small period was done to accommodate for the smaller period of 635 days of available weather and leachate inflow data in contrast to the longer period where only water inflow data was available.

For evaluation of the forecast quality, we use the \ac{MAPE} metric defined as follows for multi-step forecasts:

\begin{equation}
\text{MAPE} = \frac{100}{M*N} \sum_{i=0}^{M-1}  \sum_{j=1}^{N} | \frac{y_{i+j} - \hat{y}_{i+j}}{y_{i+j}} |
\end{equation}
where $M$ is the number of forecasts, $N$ number of forecast steps in individual forecast, $y_x$ is the true value when looking $x$ time-steps into the future and $\hat{y_x}$ is the respective predictor output. 

\subsection{Energy saving potential evaluation}\label{savings_calc}
Using the dataset of the data on energy consumption within the Micro-Grid of two neighboring companies, the following formula is used to calculate short-term forecast-based $\mathrm{CO}_\mathrm{2}$ savings $E_s$:

\begin{equation}
E_s =   \underbrace{\sum_{i=1}^{365}  C(x_i)}_\text{annual energy consumption}  -  \underbrace{
	0.5 \times \sum_{i=1}^{52} C(x_i)
}_\text{peak shift}
\end{equation}
where $x_i$ is the pumping volume in m³ at day $i$. The calculation is done on the basis of the assumption that a weekly energy intensive activity is being done at the facility that takes up half a day.

The mid-term energy savings $E_m$ are calculated as follows:
\begin{equation}
E_m = \sum_{j=1}^{j=52} \min(\underbrace{\sum_{i \in W_j}^{}  C(x_i)}_\text{workday energy} , \underbrace{
	\sum_{i \notin K_j}^{}  C(260-x_i)
}_\text{weekend energy shift capacity}
)
\end{equation}
where the sets $W_j$ and $K_j$ contain the working days  (Mon-Fr) and weekend days (Sat, Sun) of the week $j$, respectively. In the right-hand term, the maximal pumping capacity of 260 $m^3$ is subtracted from the individual historical pumping. The shifting to non-working day is motivated by peak times in energy demand when the landfill is operated during working days apparent when looking at the total energy consumption of the facility.

Additionally, the forecast uncertainty expressed via the \ac{MAPE} metric can be factored into the savings assuming that shifts will only be made up to the level that the forecast can be trusted. Utilizing the associativity of the $\min$ function, the following formula is used:

\begin{equation}
E_m' = (1-\text{MAPE} ) * E_m
\end{equation}

Based on the assumption that the shifted demand can be fullfilled using renewable energy production, all calculations of \mbox{$\mathrm{CO}_\mathrm{2}$-savings} use the figure of 523 g/kWh of \mbox{$\mathrm{CO}_\mathrm{2}$-emmissions}. This figure is based on the German energy production average in 2016.\cite{Icha.2019}

\begin{equation}
\mathrm{CO}_\mathrm{2}(E) =    \underbrace{.523 \frac{\text{kg}}{\text{Wh}}}_\text{CO2-factor} \times \underbrace{E}_\text{energy savings}
\end{equation}

\section{Results}\label{Results}
The evaluation of the different forecasting models presented in ~\ref{forecasting_models} resulted in the following forecast \ac{MAPE} depending on forecast length:

\begin{table}[htbp]
	\caption{\ac{MAPE} of the forecasting algorithms (lower is better). Trained on the datasets described in \ref{datasets}, Grid search was applied to find the best parameters for parameterized models. Note that the Boosting algorithm and attention \ac{RNN} could not forecast multiple days.}
	\begin{center}
		\begin{tabular}{|c|c| c | c| c|}
			\hline
			\textbf{Model} & \textbf{\textit{Dataset}}&\textbf{\textit{1 day}} & \textbf{\textit{7 days}} & \textbf{\textit{30 days}} \\
			\hline
			Baseline & inflow & 13.31  & 26.74  & 39.87\\
			ARIMA& inflow & 15.56& 29.95 & 47.56\\
			\ac{OLS} regression & merged& 12.76 & 24.71 & 33.00\\
			\ac{OLS} regression & inflow & \textbf{11.54} & 21.73 & \textbf{29.59}\\
			Attention RNN & merged & 15.35 & n/a & n/a\\
			Attention RNN & inflow & 15.26 & n/a & n/a\\
			Boosting & inflow & 11.78  & n/a & n/a \\
			\ac{ANN} & merged &  11.64  & 19.52& 29.90\\
			\ac{ANN} & inflow & 12.01  & \textbf{19.38}  & 29.90 \\
			\hline
		\end{tabular}
		\label{mape}
	\end{center}
\end{table}

For the baseline, $K$ values of 1,1, and 30 provided the best MAPE for 1,7, and 30-day forecasts respectively

For \ac{ANN}, the best results were achieved using a linear activation function, and a single hidden layer with dropout regularization and a dropout rate of 0.2. The training was done in mini batches with batch size 20 and ADAM.

These results have been used to optimize energy consumption in two of the three described scenarios~\ref{scenarios} in Table~\ref{savings_potential_results} for the historical pumping data in 2018.

\begin{table}[htbp]
	\caption{Potential peak load savings depending on forecast length calculated using the formulas in \ref{savings_calc}.}
	\begin{center}
		\begin{tabular}{|c|c|c|}
			\hline
			\textbf{Forecast length} & \textbf{Annual  savings in \%}& \textbf{$\mathrm{CO}_\mathrm{2}$ in kg}\\
			\hline
			short-term $E_s$ & 4.75 & 632\\
			mid-term $E_m$ & 28.79 & 3834 \\
			mid-term $E_m'$ (inc. MAPE) & 23.21 & 3091 \\
			\hline
		\end{tabular}
	\end{center}
	\label{savings_potential_results}
\end{table}



\section{Discussion}\label{Discussion}

\subsection{Forecast quality}

The given prediction quality indicates that the expected non-linear relationship between the weather data is complex as the exogeneous data did not provide a significant improvement in prediction quality. This nonexisting increase in performance can have a multitude of reasons such as sensory inaccuracy, poor data quality, or other external factors not measured.

In comparison, shallow models provided a lower (better) \ac{MAPE} score. Analyzing the parameters during grid search indicated that increasing model size by adding layers to the \ac{ANN} and changing the activation function from linear to a rectified linear unit (ReLU) did decrease performance, which may indicate overfitting on the problem parameters. One advantage of shallower models is their higher efficiency, which would reduce the footprint of the forecasting itself.

This paper used classical state-of-the-art time-series methods for energy consumption methods forecasting, leaving other, holistic methods unexplored \cite{Mosavi.2019, suganthi2012energy}.

\subsection{Energy saving potential}
The energy savings presented in chapter~\ref{savings_potential_results} show that energy demand prediction for landfills offers the potential to realize significant savings. 

Calculation of the savings happened under various assumptions, for example that the energy exchange can happen freely between generators and users of energy within the microgrid without considering transmission losses . Also, it was assumed that the energy consumption of the water pumps follows a linear relationship of input power to m³ pumped. In addition, amortized costs of building renewable energy infrastructure were not looked into. 

The numbers calculating $\mathrm{CO}_\mathrm{2}$ savings assume that 100\% of the non-peak energy can be used from renewable energy sources while peaks in demand have to be fulfilled using power from the German electrical grid. While being close to reality in our use case, other leachate plants may not be subject to such assumptions. Therefore, the application of the results depends strongly on the spread in $\mathrm{CO}_\mathrm{2}$-footprint of the energy produced in the microgrid and the German energy mix. 

Future work could pair the demand forecasting algorithm with an algorithm forecasting renewable energy supply to create holistic approaches in energy optimization for the case \cite{kleissl2013solar}.
\subsection{Long-term forecast usage}
As production planning accuracy decreases in longer-term forecasts, using the longer term leachate forecast does not bear as much potential for energy planning. However, due to the nature of the minimum and maximum water storage requirements described in~\ref{intro}, long-term forecast accuracy can help in managing the leachate treatment. As the climate shifts towards more extreme weather conditions with climate change, forecasting has the potential to assist proactive adaption pumping volume to periods of drought and flood \cite{field2012managing}.

\section{Conclusion}\label{Conclusion}
Analyzing a data set of leachate processing in an open landfill within a Microgrid, this paper evaluated state-of-the-art machine learning models for their effectiveness in energy usage forecasting of open landfill leachate pumps.

Shallow models showed the lowest \ac{MAPE}, significantly outperforming a persistence baseline for both, long-term (30 days), mid-term (7 days) and short-term (1 day) forecast. Using mid-term forecasts, a potential decrease of up to 23\% in peak energy demand was found that could lead to a reduction of up to 3,091 kg in $\mathrm{CO}_\mathrm{2}$-emissions per year when shifting peak load within an existing Microgrid.

The described methods require only low financial investments for energy-management hardware for training and deployment, making them suitable for usage in \acp{SME}.

\begin{acronym}[Bash]
	\acro{MAPE}{Mean Absolute Percentage Error}
	\acro{ARIMA}{Autoregressive Integrated Moving Average}
	\acro{DSM}{Demand Side Management}
	\acro{ANN}{Artificial Neural Network}
	\acro{OLS}{Ordinary least squares}
	\acro{RNN}{Recurrent Neural Network}
	\acro{dena}{German Energy Agency}
	\acro{DER}{Distributed Energy Ressource}
	\acro{PCC}{Point of Common Coupling}
	\acro{DG}{Distributed Generation}
	\acro{ESS}{Energy  Storage System}
	\acro{SME}{Small and Medium sized Enterprise}
\end{acronym}

\bibliographystyle{IEEEtran}
\bibliography{EWIMA}

\begin{thebibliography}{10}
\providecommand{\url}[1]{#1}
\csname url@samestyle\endcsname
\providecommand{\newblock}{\relax}
\providecommand{\bibinfo}[2]{#2}
\providecommand{\BIBentrySTDinterwordspacing}{\spaceskip=0pt\relax}
\providecommand{\BIBentryALTinterwordstretchfactor}{4}
\providecommand{\BIBentryALTinterwordspacing}{\spaceskip=\fontdimen2\font plus
\BIBentryALTinterwordstretchfactor\fontdimen3\font minus
  \fontdimen4\font\relax}
\providecommand{\BIBforeignlanguage}[2]{{%
\expandafter\ifx\csname l@#1\endcsname\relax
\typeout{** WARNING: IEEEtran.bst: No hyphenation pattern has been}%
\typeout{** loaded for the language `#1'. Using the pattern for}%
\typeout{** the default language instead.}%
\else
\language=\csname l@#1\endcsname
\fi
#2}}
\providecommand{\BIBdecl}{\relax}
\BIBdecl

\bibitem{Tegart.1990}
W.~J.~M. Tegart, G.~Sheldon, and D.~C. Griffiths, \emph{Climate change: The
  IPCC impacts assessment}.\hskip 1em plus 0.5em minus 0.4em\relax Canberra:
  {Australian Govt. Pub. Service}, 1990.

\bibitem{Pachauri.2015}
R.~K. Pachauri and L.~Mayer, Eds., \emph{Climate change 2014: Synthesis
  report}.\hskip 1em plus 0.5em minus 0.4em\relax Geneva, Switzerland:
  {Intergovernmental Panel on Climate Change}, 2015.

\bibitem{PROGRAMME.2019}
U.~N.~E. PROGRAMME, \emph{EMISSIONS GAP REPORT 2018}.\hskip 1em plus 0.5em
  minus 0.4em\relax [S.l.]: UNEP, 2019.

\bibitem{BundesministeriumfurUmweltNaturschutzundnukleareSicherheit.Juni2018}
\BIBentryALTinterwordspacing
{Bundesministerium f{\"u}r Umwelt, Naturschutz und nukleare Sicherheit},
  ``Klimaschutzbericht 2017: Zum aktionsprogramm klimaschutz 2020 der
  bundesregierung,'' Berlin. [Online]. Available:
  \url{www.bmu.de/publikationen}
\BIBentrySTDinterwordspacing

\bibitem{Jacobsson.2006}
S.~Jacobsson and V.~Lauber, ``The politics and policy of energy system
  transformation---explaining the german diffusion of renewable energy
  technology,'' \emph{Energy Policy}, vol.~34, no.~3, pp. 256--276, 2006.

\bibitem{Papaefthymiou.2016}
G.~Papaefthymiou and K.~Dragoon, ``Towards 100{\%} renewable energy systems:
  Uncapping power system flexibility,'' \emph{Energy Policy}, vol.~92, pp.
  69--82, 2016.

\bibitem{Mwasilu.2014}
F.~Mwasilu, J.~J. Justo, E.-K. Kim, T.~D. Do, and J.-W. Jung, ``Electric
  vehicles and smart grid interaction: A review on vehicle to grid and
  renewable energy sources integration,'' \emph{Renewable and Sustainable
  Energy Reviews}, vol.~34, pp. 501--516, 2014.

\bibitem{Olivares.2014}
D.~E. Olivares, A.~Mehrizi-Sani, A.~H. Etemadi, C.~A. Canizares, R.~Iravani,
  M.~Kazerani, A.~H. Hajimiragha, O.~Gomis-Bellmunt, M.~Saeedifard,
  R.~Palma-Behnke, G.~A. Jimenez-Estevez, and N.~D. Hatziargyriou, ``Trends in
  microgrid control,'' \emph{IEEE Transactions on Smart Grid}, vol.~5, no.~4,
  pp. 1905--1919, 2014.

\bibitem{Gungor.2013}
V.~C. Gungor, D.~Sahin, T.~Kocak, S.~Ergut, C.~Buccella, C.~Cecati, and G.~P.
  Hancke, ``A survey on smart grid potential applications and communication
  requirements,'' \emph{IEEE Transactions on Industrial Informatics}, vol.~9,
  no.~1, pp. 28--42, 2013.

\bibitem{Hatziargyriou.2731Jan.2002}
N.~Hatziargyriou, G.~Contaxis, M.~Matos, J.~Lopes, G.~Kariniotakis, D.~Mayer,
  J.~Halliday, G.~Dutton, P.~Dokopoulos, A.~Bakirtzis, J.~Stefanakis,
  A.~Gigantidou, P.~O'Donnell, D.~McCoy, M.~J. Fernandes, J.~Cotrim, and A.~P.
  Figueira, ``Energy management and control of island power systems with
  increased penetration from renewable sources,'' in \emph{2002 IEEE Power
  Engineering Society Winter Meeting. Conference Proceedings (Cat.
  No.02CH37309)}.\hskip 1em plus 0.5em minus 0.4em\relax IEEE, 27-31 Jan. 2002,
  pp. 335--339.

\bibitem{Strbac.2008}
G.~Strbac, ``Demand side management: Benefits and challenges,'' \emph{Energy
  Policy}, vol.~36, no.~12, pp. 4419--4426, 2008.

\bibitem{Seidl.062016}
\BIBentryALTinterwordspacing
H.~Seidl, C.~Schenuit, and M.~Teichmann, ``Roadmap demand side management:
  Industrielles lastmanagement f{\"u}r ein zukunftsf{\"a}higes energiesystem.
  schlussfolgerungenaus dem pilotprojekt dsm bayern.'' Berlin. [Online].
  Available:
  \url{https://www.dena.de/themen-projekte/energiesysteme/flexibilitaet-und-speicher/demand-side-management/}
\BIBentrySTDinterwordspacing

\bibitem{S.Renou.2008}
\BIBentryALTinterwordspacing
{S. Renou}, {J.G. Givaudan}, {S. Poulain}, {F. Dirassouyan}, and {P. Moulin},
  ``Landfill leachate treatment: Review and opportunity,'' \emph{Journal of
  Hazardous Materials}, vol. 150, no.~3, pp. 468--493, 2008. [Online].
  Available:
  \url{http://www.sciencedirect.com/science/article/pii/S0304389407013593}
\BIBentrySTDinterwordspacing

\bibitem{Smith.2019}
\BIBentryALTinterwordspacing
T.~G. Smith, ``pmdarima: Arima estimators for python,'' 2019. [Online].
  Available: \url{https://www.alkaline-ml.com/pmdarima/index.html}
\BIBentrySTDinterwordspacing

\bibitem{Qin.472017}
\BIBentryALTinterwordspacing
Y.~Qin, D.~Song, H.~Chen, W.~Cheng, G.~Jiang, and G.~Cottrell, ``A dual-stage
  attention-based recurrent neural network for time series prediction.''
  [Online]. Available: \url{http://arxiv.org/pdf/1704.02971v4}
\BIBentrySTDinterwordspacing

\bibitem{ke2017lightgbm}
G.~Ke, Q.~Meng, T.~Finley, T.~Wang, W.~Chen, W.~Ma, Q.~Ye, and T.-Y. Liu,
  ``Lightgbm: A highly efficient gradient boosting decision tree,'' in
  \emph{Advances in Neural Information Processing Systems}, 2017, pp.
  3146--3154.

\bibitem{Icha.2019}
\BIBentryALTinterwordspacing
P.~Icha, ``Entwicklung der spezifischen kohlendioxid-emissionen des deutschen
  strommix in den jahren 1990 - 2018,'' Dessau-Ro{\ss}lau. [Online]. Available:
  \url{https://www.umweltbundesamt.de/sites/default/files/medien/1410/publikationen/2019-04-10_cc_10-2019_strommix_2019.pdf}
\BIBentrySTDinterwordspacing

\bibitem{Mosavi.2019}
A.~Mosavi, M.~Salimi, S.~{Faizollahzadeh Ardabili}, T.~Rabczuk,
  S.~Shamshirband, and A.~R. Varkonyi-Koczy, ``State of the art of machine
  learning models in energy systems, a systematic review,'' \emph{Energies},
  vol.~12, no.~7, p. 1301, 2019.

\bibitem{suganthi2012energy}
L.~Suganthi and A.~A. Samuel, ``Energy models for demand forecasting---a
  review,'' \emph{Renewable and Sustainable Energy Reviews}, vol.~16, no.~2,
  pp. 1223--1240, 2012.

\bibitem{kleissl2013solar}
J.~Kleissl, \emph{Solar energy forecasting and resource assessment}.\hskip 1em
  plus 0.5em minus 0.4em\relax {Academic Press}, 2013.

\bibitem{field2012managing}
C.~B. Field, V.~Barros, T.~F. Stocker, and Q.~Dahe, \emph{Managing the risks of
  extreme events and disasters to advance climate change adaptation: special
  report of the intergovernmental panel on climate change}.\hskip 1em plus
  0.5em minus 0.4em\relax {Cambridge University Press}, 2012.

\end{thebibliography}


\begin{thebibliography}{1}

\bibitem{GeschaftsstellederArbeitsgruppeErneuerbareEnergienStatistikAGEEStatamUmweltbundesamt.2019}
{Gesch{\"a}ftsstelle der Arbeitsgruppe Erneuerbare Energien-Statistik
  (AGEE-Stat) am Umweltbundesamt}, ``Erneuerbare energien in deutschland 2018:
  Daten zur entwicklung im jahr 2018,'' 2019.

\end{thebibliography}

\end{document}